\documentclass[preprint, 12pt]{elsarticle}

\usepackage{amssymb}
\usepackage{amsmath}
\usepackage{tabularx}
\usepackage{hyperref}
\usepackage{xcolor}
\usepackage{geometry}
\usepackage{multirow}
\usepackage{ragged2e}
\newcolumntype{L}{>{\RaggedRight\arraybackslash}X}
\usepackage{pdfpages}
\usepackage{booktabs}
\newcommand{\tabitem}{~~\llap{\textbullet}~~}
\usepackage{ragged2e}   
\newcolumntype{Y}{>{\RaggedRight\arraybackslash}X}  
\newcolumntype{P}{>{\justifying\arraybackslash}X}

\journal{Maturitas}

\begin{document}

\begin{frontmatter}



\author[label1]{Bram M.A. van Dijk}
\author[label1]{Armel E.J.L. Lefebvre}
\author[label1,label2]{Marco R. Spruit\corref{cor1}}

\affiliation[label1]{organization={Leiden University Medical Center, Department of Public Health and Primary Care},
             addressline={Turfmarkt 99},
             postcode={2511 DP},
             city={The Hague},
             country={The Netherlands}}
\affiliation[label2]{organization={Leiden University, Leiden Institute of Advanced Computer Science},
             addressline={Einsteinweg 55},
             postcode={2333 CC},
             city={Leiden},
             country={The Netherlands}}
\ead{m.r.spruit@lumc.nl}
\cortext[cor1]{Corresponding author at: Health Campus The Hague, Turfmarkt 99, 2511 DP The Hague.}

\title{Welzijn.AI: Developing Responsible Conversational AI for Elderly Care through Stakeholder Involvement}




\begin{abstract}

\paragraph{Objectives} We present \texttt{Welzijn.AI} as new digital solution for monitoring (mental) well-being in elderly populations, and illustrate how development of systems like \texttt{Welzijn.AI} can align with guidelines on responsible AI development. 

\paragraph{Study design} Three evaluations with different stakeholders were designed to disclose new perspectives on the strengths, weaknesses, design characteristics, and value requirements of \texttt{Welzijn.AI}. Evaluations concerned expert panels and involved patient federations, general practitioners, researchers, and the elderly themselves. Panels concerned interviews, a co-creation session, and feedback on a proof-of-concept implementation. 

\paragraph{Main outcome measures}
Interview results were summarized in terms of \texttt{Welzijn.AI}'s strengths, weaknesses, opportunities and threats. The co-creation session ranked a variety of value requirements of \texttt{Welzijn.AI} with the Hundred Dollar Method. User evaluation comprised analysing proportions of (dis)agreement on statements targeting \texttt{Welzijn.AI}'s design characteristics, and ranking desired social characteristics.

\paragraph{Results}
Experts in the panel interviews acknowledged \texttt{Welzijn.AI}'s potential to combat loneliness and extract patterns from elderly behaviour. The proof-of-concept evaluation complemented the design characteristics most appealing to the elderly to potentially achieve this: empathetic and varying interactions. Stakeholders also link the technology to the implementation context: \texttt{Welzijn.AI} can unlock an individual's social network, but practice sessions and continuous support should also be available to empower users. Yet, non-elderly and elderly experts also disclose challenges in properly understanding the application; non-elderly experts also highlight issues concerning privacy. 

\paragraph{Conclusions} 
Incorporating all stakeholder perspectives in system development remains challenging. Still, our results benefit researchers, policy makers, and health professionals that aim to improve elderly care with technology.

\end{abstract}




\end{frontmatter}


\section{Introduction}\label{intro}
\noindent The number of elderly individuals ($\geq$ 60 years) in the population is increasing and estimated to be one in six globally by 2030 \citep{WHO2023}. Maintaining physical and mental well-being is challenging for the elderly, who are vulnerable to social isolation \citep{WHO2021}, loss of cognitive functioning \citep{abdoli2022global}, and live in a time of labour shortages in healthcare \citep{EU2023}, meaning a smaller amount of caregivers treats more elderly with complex needs.

Artificial Intelligence (AI), in particular conversational systems driven by Large Language Models (LLMs), can support both the elderly and caregivers by enabling remote monitoring of elderly well-being. Current examples include reminding elderly of medication and appointments, and providing social interaction \citep{sapci2019innovative}. More extensive monitoring, however, could also enable elderly to live independently for longer, and early signalling of physical and mental health problems, thereby reducing pressure on healthcare systems.  Though recent work on this topic has identified challenges specific to tailoring AI to the elderly \citep{klaassen2024review, wolfe2025caregiving}, little work embeds stakeholders in the \textit{development} of AI systems for elderly care. 

\begin{figure}[t]
\centering
\includegraphics[width=\textwidth]{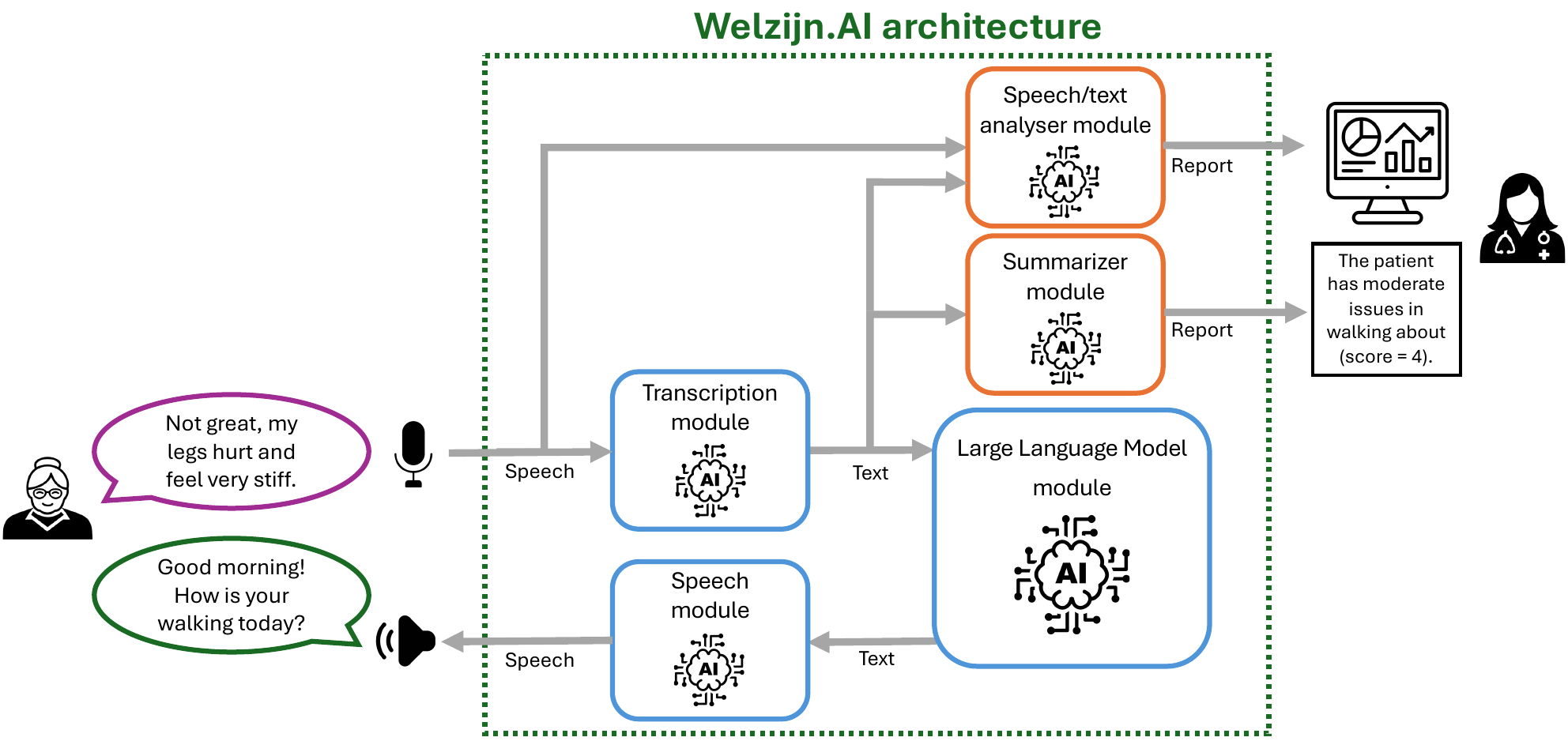}
\caption{\texttt{Welzijn.AI} comprises modules for interaction (blue) and analysis (orange). Regarding \textbf{interaction}, the \textit{transcription module} concerns an open weights Whisper model \cite{radford2023robust} that transcribes user speech to text as input to the LLM module. The \textit{Large Language Model module} supports the conversation with the open weights Llama3.3-70B model \cite{grattafiori2024llama}, prompted to structure the conversation around the EQ-5D-5L. The \textit{speech module} is an open weights CoquiTTS VITS model that converts LLM responses to natural speech \cite{kim2021conditional}. Regarding \textbf{analysis}, user input in text and audio formats are analysed in the \textit{speech/text analyser module}. Text is also input to the \textit{summarizer module}, a Llama3.3-70B model that summarizes user responses into a score.}
\label{fig:architecture}
\end{figure}

In this paper we introduce \texttt{Welzijn.AI}, a conversational AI system driven by state-of-the-art LLMs, transcription and speech generation models to retrieve clinically relevant information about elderly quality of life and well-being. \texttt{Welzijn.AI} comprises a voice-controlled chatbot that asks how an elderly individual is doing regarding mobility, self-care, daily activities, discomfort/pain, and mental health, i.e. the topics in the EQ-5D-5L \cite{brooks1996euroqol}. The system summarizes user responses for each topic into scores (1-5) as common in clinical use of the EQ-5D-5L. \texttt{Welzijn.AI} also analyses elderly language use for known language biomarkers of mental well-being such as pitch and topical coherence \citep{drougkas2024multimodal,figueroa2022automatic, spruit2022exploring}. A high-level overview of the architecture is given in Figure \ref{fig:architecture}, and provides the background against which simpler implementations for the stakeholder evaluations in this paper should be understood.     

We position \texttt{Welzijn.AI} as a \textit{use case} of AI system development in the early phases of the CEHRES roadmap for the development of digital health applications \citep{van2011holistic}. Doing so, we show how researchers can addresses common challenges in AI system development for health: such systems are rarely evaluated beyond the lab; lack alignment to established development frameworks, do not align with the context of prospective users, and fail to address a healthcare context in which medical decision making based on data and AI becomes more prominent  \citep{hagendorff2020ethics,nair2024comprehensive}.  

We report three complementary evaluations of early versions of \texttt{Welzijn.AI} with various experts/stakeholders, that provide novel insights in stakeholders' views on emerging technologies like \texttt{Welzijn.AI}. Our results show, among other things, that though most stakeholders generally acknowledge the potential of the system in analysing patterns in elderly (linguistic) behaviour, properly understanding the goal and use of the system is still challenging. Stakeholders also suggested measures to address this issue, such as demo and test sessions and a help desk. Overall, we believe that our results are useful to researchers, developers, and caregivers aiming to develop responsible AI for healthcare, that takes stakeholder perspectives into account. 

\begin{figure}[t]
\centering
\includegraphics[width=\textwidth]{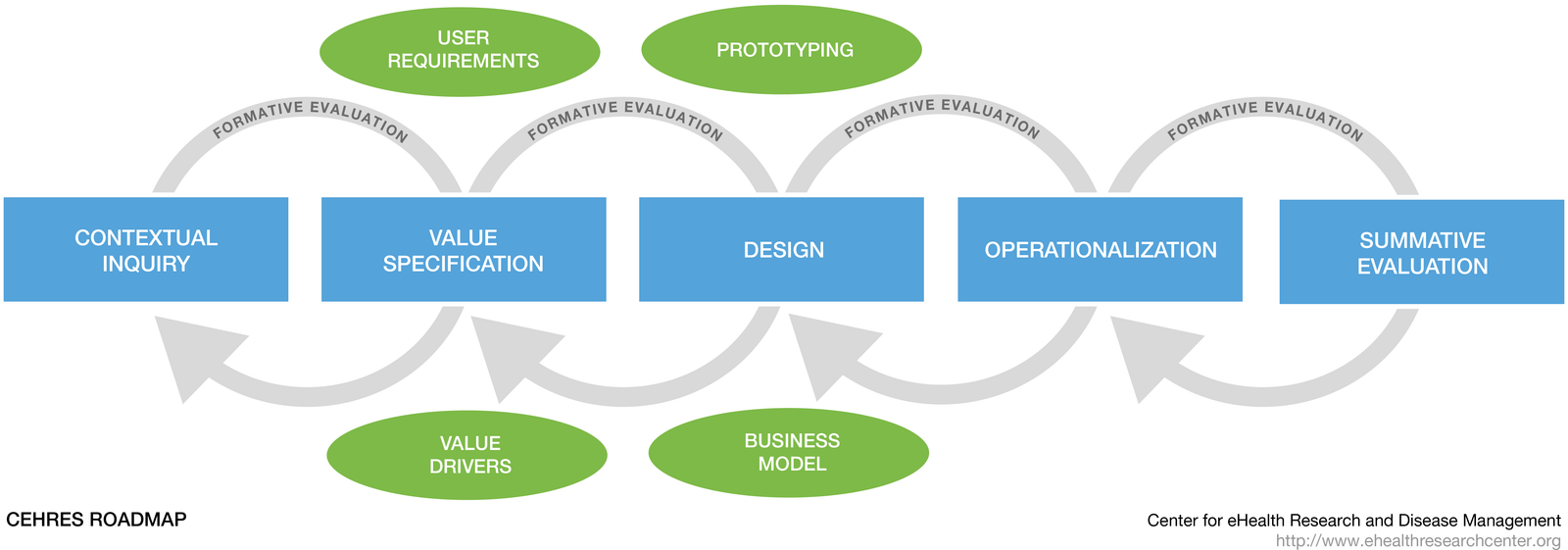}
\caption{The CEHRES roadmap (reproduced from \citep{van2011holistic}) involves five phases. In short, \textit{Contextual Inquiry} discloses the contexts of prospective stakeholders of the system; \textit{Value Specification} entails translating stakeholder values into user requirements; \textit{Design} involves the (co-)creation of prototypes with stakeholders; \textit{Operationalisation} involves bringing the technology to the market, and the \textit{Summative Evaluation} concerns an evaluation of the technology's actual use. In this work we focus mainly on the first two phases.}
\label{fig:roadmap}
\end{figure}

\section{Methods}\label{methods}
\noindent Following the \textit{contextual inquiry} and \textit{value specification} phases of the CEHRES roadmap, we conducted three stakeholder evaluations to gain deeper understanding of prospective stakeholders and their contexts, further described below. As the authors of CEHRES note, the development of systems like \texttt{Welzijn.AI} is not a linear but a dynamic, iterative process where stakeholder evaluation occurs in multiple phases in development. Especially when a technology is interactive, stakeholders should provide feedback in multiple iterations to better understand their contexts and perspectives \cite{van2011holistic}.

These iterations create a tension; from a technology development perspective, small and quick stakeholder evaluations are preferable, as many follow-up evaluations are likely required. An example concerns an expert panel that quickly provides feedback on a specific aspect of a system, like the monitoring function of \texttt{Welzijn.AI}. From a research perspective however, experiments are often the reference point, which commonly involves a large and diverse sample, ethical review, protocols, data management plans, informed consent, etc., which are less suitable to quick iterations. 

The stakeholder evaluations in this paper are better understood as \textit{expert panels} from a technology development perspective rather than experiments. Experts were selected because of their professional knowledge and experience as is common in software engineering \citep[e.g. as in][]{beecham2005using}, or because they represent the target audience and implementation context well. In this paper, experts comprised GPs, (assistant) professors elderly care, software engineers, representatives of patient federations, and elderly individuals in nursing homes. All experts received disclosure regarding the goals of the panel, verbally consented to participation, could withdraw from participation at any moment, and no sensitive nor identifying information was asked, as common in experiments, but no separate ethical review took place. In the next sections we describe each panel in more detail.          

\subsection{Expert panel interviews} \noindent Here the aim was to gain deeper understanding of the elderly care context in which stakeholders operate, following the \textit{contextual inquiry} phase in CEHRES. Six Dutch experts were involved: a GP, two representatives of an association of Dutch GPs, a clinical psychologist, a representative of the Dutch patient federation, and an assistant professor intergenerational care. Stakeholders were interviewed individually about the question how technology like \texttt{Welzijn.AI} can support vulnerable elderly. Interviews were semi-structured. First, the goal of \texttt{Welzijn.AI} was described, as conversational system that via analysis of language monitors (mental) well-being, allowing caregivers to take action when well-being is declining. Thereafter, experts were asked questions about the development of the technology, regarding its general purpose, practical use, opportunities, and threats. Interviews were manually examined and results were summarised in a Strengths, Weaknesses, Opportunities, and Threats (SWOT) model that is discussed in Section \ref{results:SWOT}. 

\subsection{Expert co-creation session} \noindent This session comprised stakeholder evaluations of \texttt{Welzijn.AI}'s monitoring and signalling function in its broadest sense: besides tracking (mental) well-being, the system was envisioned to also signal longer-than-usual periods without interaction. A representative of the Dutch association for informal caregivers and a robotics student represented two informal caregivers, a geriatrics specialist represented a professional caregiver, and a computer scientist represented a developer. The goal was to identify specific stakeholder values and value requirements to implement these values, following the \textit{value specification} phase of the CEHRES roadmap.

Each stakeholder type (developer, professional caregiver, informal caregiver) first created a use case about an elderly individual and \texttt{Welzijn.AI}'s monitoring function. Each stakeholder then distributed 100 dollars over the use cases, i.e. ranked them with the Hundred Dollar Method (HDM) \citep{leffingwell2000managing}. For the use case with most dollars, core values were extracted by all stakeholders. Subsequently, value requirements that establish or resolve conflicts in core values were identified. These requirements targeted three aspects of \texttt{Welzijn.AI} and were ranked individually with the HDM: \texttt{Welzijn.AI}'s underlying \textit{technology}, its immediate \textit{environment}, and the \textit{user} itself. Results are discussed in Section \ref{results:co-creation}.

\begin{figure}[t]
\centering
\includegraphics[width=\textwidth]{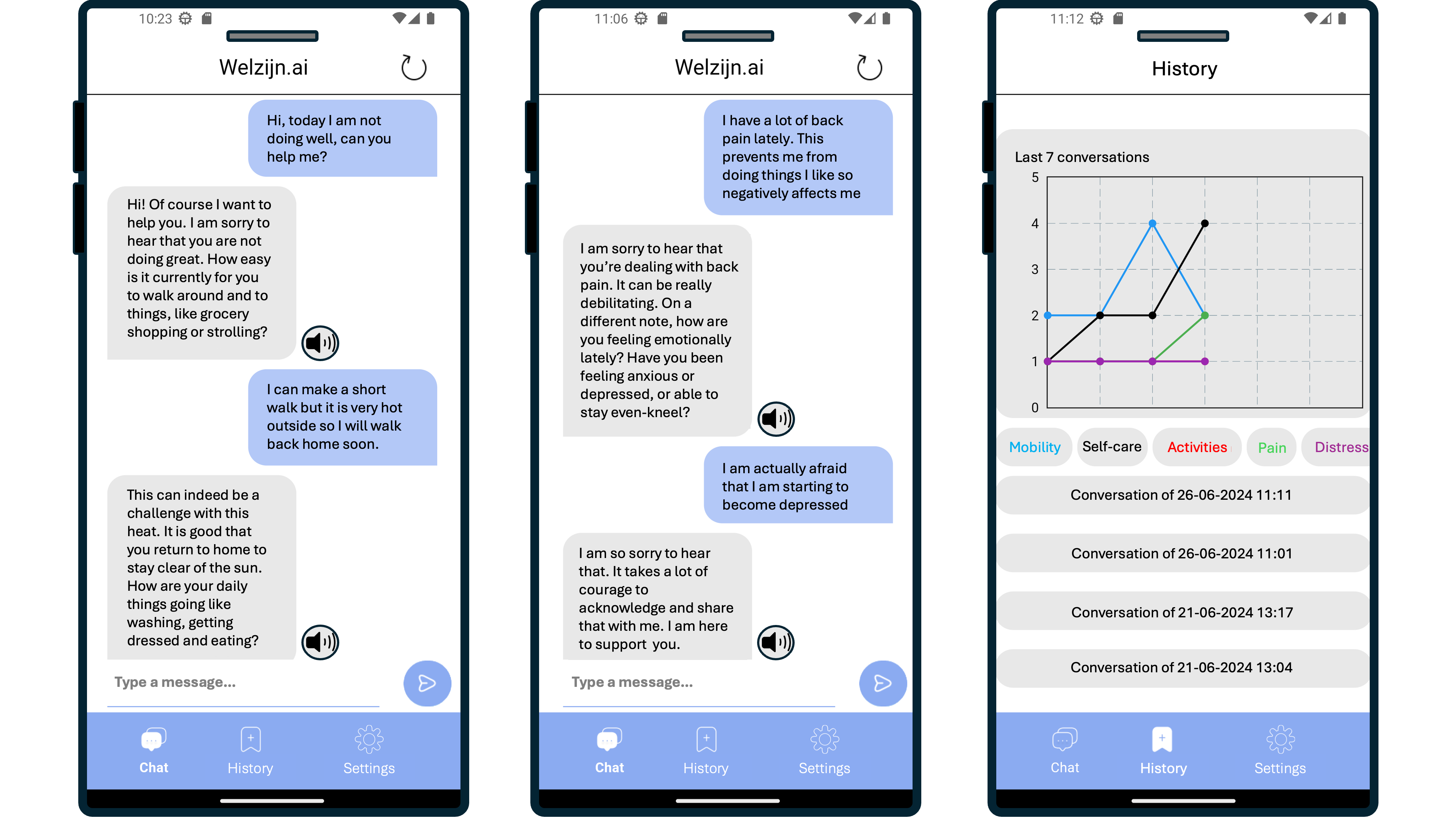}
\caption{Proof-of-concept static application interface of \texttt{Welzijn.AI}, translated to English. Left and middle panels illustrate how \texttt{Welzijn.AI} was presented to the elderly (Section \ref{elderly_eval}), with real conversations about various EQ-5D-5L topics (mobility, self-care and mood). Right panel shows a work-in-progress dashboard illustrating how text-to-score module score outputs can be tracked over multiple conversations and topics, allowing signalling decline in well-being and quality of life.}
\label{fig:proof-of-concept_ex}
\end{figure}

\begin{table}[t]
\centering
\footnotesize
\begin{tabularx}{\textwidth}{|L|L|L|}
 \hline
 \textbf{Characteristic} & \textbf{Example} & \textbf{Answer} \\
 \hline
 \multirow{4}{=}{\textbf{1. Accessibility}} 
  & \tabitem The functions of the icons on the screen are clear & \multirow{4}{=}{Strongly disagree, Disagree, Neither agree nor disagree, Agree, Strongly agree} \\
  & \tabitem The language used is easy to grasp & \\
  \hline
  \multirow{4}{=}{\textbf{2. Comprehensibility}} 
  & \tabitem The system looks needlessly complex & \multirow{4}{=}{Strongly disagree, Disagree, Neither agree nor disagree, Agree, Strongly agree} \\
  & \tabitem I think I can master this system reasonably fast & \\
  \hline
  \multirow{4}{=}{\textbf{3. Intention to use}} 
  & \tabitem I would recommend this system to others & \multirow{4}{=}{Strongly disagree, Disagree, Neither agree nor disagree, Agree, Strongly agree} \\
  & \tabitem I would only use this system if my GP recommends it & \\
  \hline
\multirow{4}{=}{\textbf{4. Perceived trust}} 
  & \tabitem I think this system is reliable & \multirow{4}{=}{Strongly disagree, Disagree, Neither agree nor disagree, Agree, Strongly agree} \\
  & \tabitem Overall I have trust in this system & \\
  \hline
  \multirow{2}{=}{\textbf{5. Satisfaction}} 
  & \tabitem Frustrating - Satisfying & \multirow{2}{=}{ O O O O O } \\
  & \tabitem Boring - Fun &  \\
  \hline
  \multirow{2}{=}{\textbf{6. Human-likeness}} 
  & \tabitem Human-like - Machine-like & \multirow{2}{=}{ O O O O O } \\
  & \tabitem Fake - Real & \\
\hline
\end{tabularx}
\caption{Example items of the expert proof-of-concept evaluation and their answer options. For accessibility, comprehensibility, intention to use, and perceived trust, \textit{statements} were used, and elderly responded with (dis)agreement on a 5-point Likert scale. Satisfaction and human-likeness of the system were measured with \textit{semantic differential scales}, where elderly chose one point on a 5-point scale.}
\label{tab:expert_POC_items}
\end{table}

\subsection{Expert proof-of-concept evaluation} \noindent This evaluation presented a proof-of-concept static interface (Figure \ref{fig:proof-of-concept_ex}) to an expert panel consisting of 20 elderly individuals $(\bar{x}_{age} = 83.2, \sigma_{age} = 8.1)$ from the south-western part of The Netherlands, to identify 1) their user perspective and 2) \texttt{Welzijn.AI}'s desired social characteristics. The goal was to obtain opinions on an initial design to inform further development, fitting the \textit{contextual inquiry} phase of the CEHRES roadmap.

\textbf{User perception} was operationalised with a set of items \footnote{All questions can be found in the supplementary materials.} that targeted six characteristics linked to the use of a system in the literature: \textit{accessibility} \citep{diaz2014accessibility}, \textit{comprehensibility} \citep{davis1989perceived}, \textit{intention to use} \citep{heerink2008influence}, \textit{perceived trust} \citep{meng2022trust}, \textit{satisfaction} \citep{gelderman1998relation}, and \textit{human-likeness} \citep{li2021machinelike}. See Table \ref{tab:expert_POC_items} for an illustration of items and answers.

The analysis focused on the proportion of positively evaluated items for each characteristic. For items concerning \textit{statements} (items 1 through 4 in Table \ref{tab:expert_POC_items}), only (strong) agreement (or disagreement in case of negatively formulated statements) indicated a positive perception regarding a characteristic. For the items concerning \textit{semantic differential scales} (items 5 and 6 in Table \ref{tab:expert_POC_items}), only the two levels closest to the positive end of a negative-positive pair indicated a positive perception regarding a characteristic.

\textbf{Desired social characteristics} for interactive AI systems were adapted from work by \cite{heerink2008influence, fong2003survey} and the elderly ranked them individually. Social characteristics were: 1. responding empathetic (i.e. recognise and express emotions); 2. establishing social relationships (e.g. properly greeting the user); 3. using natural cues in interaction (e.g. use of fillers, emoticons); 4. exhibiting a distinctive personality/character (e.g. chatbot tells something about itself); 5. developed social competencies (e.g. capable of small talk); 6. building a relation of trust with the user (i.e. supporting the user); 7. behaving transparently (e.g. admitting errors). The rank for each characteristic was converted to a score (7 points for rank 1, 6 for rank 2, etc.), and scores of all users were summed into a final ranking.

\section{Results}\label{results}
\subsection{Expert panel interviews}\label{results:SWOT}

\noindent Expert panel opinions are summarised in Table \ref{tab:SWOT}, and discussed in more detail below. 

\textbf{Strengths --} The majority of experts aligned on the potential of \texttt{Welzijn.AI} in reducing loneliness, and many also agreed on the valuable insights that could be obtained through the data in an individual's behaviour, which could warrant also signalling and follow-up action based on the data.  

\begin{table}[t]
\centering
\scriptsize
\begin{tabularx}{\textwidth}{|l|LLLLL|}
 \hline
 & \textbf{GP} & \textbf{PF} & \textbf{CP} & \textbf{AGP} & \textbf{AP} \\
\hline
 \textbf{S} & Tracking elderly behaviour, flagging wellbeing issues, reducing loneliness & Signalling function, link to informal caregivers, reducing loneliness & Pleasant interaction, reduction of loneliness & Insight in elderly perspectives on/use of technology through early involvement & Knowledge on individual behaviour patterns, reducing loneliness  \\
\hline
 \textbf{W} & Unclear relation to existing regulation and privacy, unclear added value for elderly, signalling function underdeveloped & Utility not clear, signalling deviations presumes knowing an individual well, app does not know physical environment & Goal unclear, lack of validation of collected medical data  & Added value unclear, unclear when a signal needs follow-up & Small difference with existing technologies, offering yet another solution instead of recognizing a problem  \\
\hline
 \textbf{O} & Mediator between home care/nursing home  & Providing guidance for simple daily tasks & Support for multiple languages, mapping social context, reaching care-avoiders  & Integration with other health systems, activating social network of elderly person that knows the person well & Embedding in plush toy for triggering care response and improving quality of life  \\
\hline
 \textbf{T} & Privacy regarding continuous monitoring  & Privacy, wrong conclusions from the data, seeing technology as replacing humans & Accessibility, information access third parties & Small malfunctions cause loss of trust  & Seeing technology as replacing humans, creating technology dependence \\
\hline
\end{tabularx}
\caption{Strengths (S), Weaknesses (W), Opportunities (O) and Threats (T) of \texttt{Welzijn.AI} as new technology in elderly care, from the perspective of a General Practitioner (GP), the Dutch Patient Federation (PF), a Clinical Psychologist (CP), an Association of Dutch GPs (AGP), and an Assistant Professor intergenerational care (AP).}
\label{tab:SWOT}
\end{table}

\textbf{Weaknesses --} Experts pointed out unclarity regarding the overall utility and goal of the \texttt{Welzijn.AI} from a user perspective (``what's in it for me''), and more broadly the need to articulate in what way this technology is different from what already exists. Two experts took issue with the output of the technology and its potential signalling function: for example, in monitoring speech, the issue how to decide what constitutes a deviation from normal patterns, and what the follow-up on a warning signal comprises (e.g. should a professional caregiver, GP, informal caregiver, or family member be warned).

\textbf{Opportunities --} Two experts emphasised the potential of \texttt{Welzijn.AI} to map and activate the social network of an elderly individual, that may know the individual well and can best decide whether further action is needed (though this presumes the network is known). Two experts further noted the potential impact on quality of life in two ways: by providing guidance through conversation in simple daily tasks, and by being embedded in a plush toy and instilling a care response and sense of purpose.        

\textbf{Threats --} For most experts, recurring issues are privacy and safety, for example when the app is used for continuous monitoring, but also the issue who has access to the data. Lastly, two experts mention the issue that \texttt{Welzijn.AI} may create the impression that technology can replace humans, i.e. it may create a dependence on technology. Another threat is that wrong conclusions are inferred from the data, because deviation from normal behaviour requires knowing what normal behaviour is.

\subsection{Expert co-creation session}\label{results:co-creation}
\begin{figure}[t]
\includegraphics[width=\textwidth]{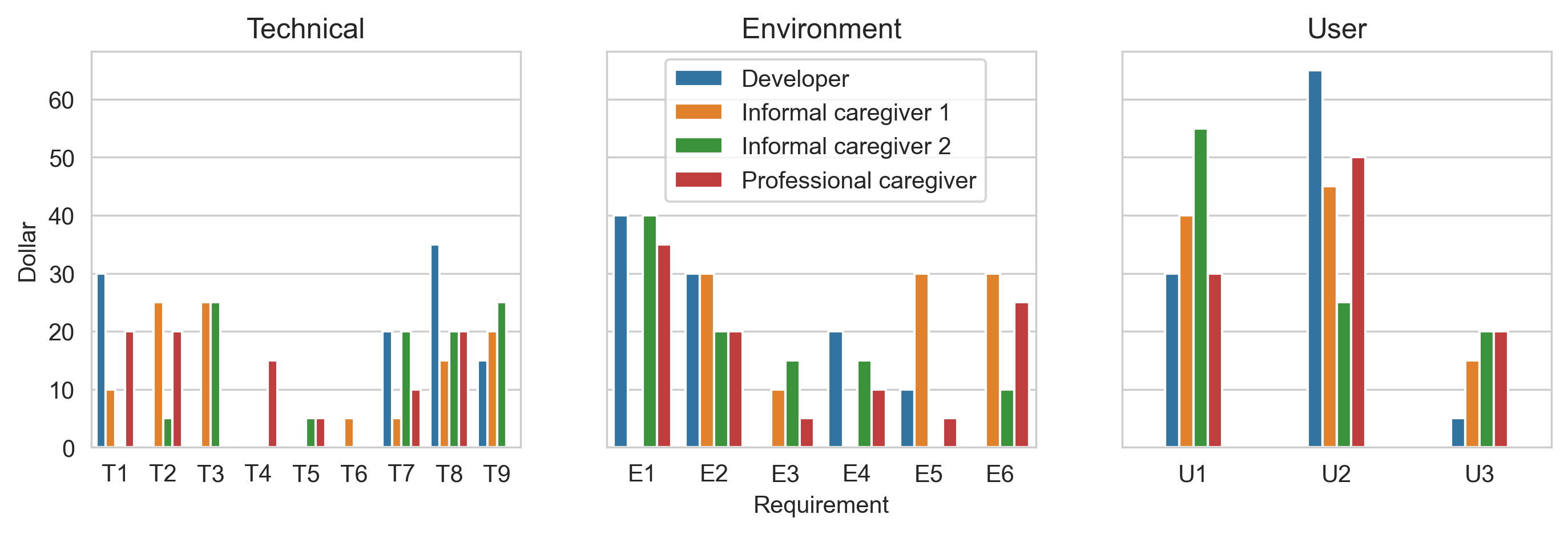}
\caption{Value requirements of \texttt{Welzijn.AI} regarding its technical, environmental, and user-related aspects, ranked with the HDM. Requirements are listed as follows. \textbf{Technical:} (T1) periodically asking consent, (T2) available on-off button, (T3) robust, portable design, (T4) automatic shutdown in case of multiple persons present, (T5) long battery life, (T6) limited human-likeness, (T7) gradual flow of the conversation, (T8) safe data storage, (T9) speech recognition. \textbf{Environment:} (E1) available WiFi, (E2) help desk, (E3) power outlet, (E4) social media functionality, (E5) promotional campaign, (E6) agreements on access to data. \textbf{User:} (U1) education for caregivers and user, (U2) demo, test and practice session, (U3) creating awareness to reduce stigma on using assistive technology.}
\label{fig:cocreate}
\end{figure}

\noindent The use case allocated most dollars concerned the scenario in which an elderly woman interacts daily with \texttt{Welzijn.AI} at a fixed time after a stroll.\footnote{All use cases can be found in the supplementary materials.} One day she does this much later because she was afraid to use the stairs, causing the system to alert a third person. Stakeholders extracted consensual core values for the design of \texttt{Welzijn.AI} from this use case, and they were \textit{autonomy}, \textit{privacy}, \textit{living independently}, and \textit{consent/voluntary use}. Hereafter, value requirements were identified and allocated dollars by stakeholders individually (for an overview see Figure \ref{fig:cocreate}).  

\subsubsection{Consensus}
\noindent We first discuss value requirements that all stakeholders valued more than zero dollar, hence all agreed on. These were a gradual flow of the conversation (T7), safe data storage (T8), help desk (E2), and demo, test and practice session (U2).\footnote{As the number of \textit{user} value requirements is small, consensus is more likely, so the requirement with the highest aggregate ranking (U2) was chosen as consensual one.} Safe data storage links to \textit{privacy} as core value; a demo, test and practice session empowers user \textit{autonomy} (esp. in contexts where users \textit{live independently}); a gradual flow of the conversation in the design and a help desk for \texttt{Welzijn.AI} may also support \textit{autonomy} and \textit{living independently}.

\subsubsection{Disparity}
\noindent We also discuss some (but not all) disparity in dollar allocation for different stakeholders. 

For example, the two \textbf{informal caregivers} diverge on various value requirements, such as periodically asking consent (T1) (contributes to \textit{consent/voluntary use}) and a promotional campaign for \texttt{Welzijn.AI} (E5) (less clear relation to specific core values), but align on robust, portable design (T3) (contributes to \textit{living independently}), the presence of speech recognition (T9) (contributes to \textit{autonomy} and \textit{living independently}) and education for caregivers and users (U1) (contributes to \textit{autonomy} and \textit{living independently}). Informal caregivers seem to have a preference for value requirements that contribute to \textit{autonomy} and \textit{living independently}.   

The \textbf{developer} allocated some value requirements zero dollar that the professional and/or informal caregiver(s) allocated much more: for example the on-off button (T2) (contributes to \textit{autonomy, consent/voluntary use, privacy}), a robust, portable design (T3) (contributes to \textit{living independently}), and agreements on access to data (E6) (contributes to \textit{privacy}). The opposite was the case for technical requirements like safe data storage (T8) (contributes to \textit{privacy}) and periodically asking consent (T1) (contributes to \textit{consent/voluntary use}). Overall, the developer distributed dollars over a smaller number of value requirements compared to other stakeholders, so prioritises more.

The \textbf{professional caregiver}, unlike other stakeholders, allocated dollars to the automatic shutdown in case of multiple persons present (T4) (contributes to \textit{privacy}), but allocated zero dollar to a robust, portable design (T3) (contributes to \textit{autonomy} and \textit{living independently}), which both informal caregivers gave much more. The professional caregiver often aligned with one or more informal caregivers, for example on agreements on access to data (E6) (contributes to \textit{privacy}), and demo, test and practice session (U2) (contributes to \textit{autonomy} and \textit{living independently}).  

\subsection{Expert proof-of-concept evaluation}\label{elderly_eval}
\noindent Here we discuss 1) opinions of elderly individuals on \texttt{Welzijn.AI} and 2) their ranking of \texttt{Welzijn.AI}’s desired social characteristics.

\subsubsection{User perception}
\noindent Results of the user perception evaluation are given in Table \ref{tab:user_eval}. Here we discuss some outcomes on specific items in the evaluation, but not all. 

Regarding \textbf{accessibility}, 64\% of the elderly responses evaluated the proof-of-concept version as accessible. Elderly were positive about font size, contrast and language use, but negative about understanding the meaning and use of the icons. Concerning \textbf{perceived trust}, 59\% of the responses indicated trust in the system. Elderly rated the system as reliable, and trusted the system's output, although they also thought that a human in the loop was required for proper functioning. For \textbf{human-likeness}, 54\% of the responses evaluated the system as human-like. Elderly experienced the system as natural, and more human than machine-like, but did not associate the system with consciousness. Concerning \textbf{intention to use}, 52\% of the responses showed intention to use the app. The elderly planned to use the app also without recommendation by a doctor, but disagreed recommending it to others. Regarding \textbf{comprehensibility}, 44\% of the responses evaluated the app as comprehensible. Elderly perceived the app as complex, and taking time to understand, although they also thought the app was useful. For \textbf{satisfaction}, 42\% of the responses displayed satisfaction with the system. Most elderly displayed mixed feelings and indifference on whether the app was fun or hard to use.

\begin{table}[t] \small
\begin{center}
\begin{tabular}{c c c}
    \hline
    \textbf{Characteristic} & \textbf{Positive perception} & \textbf{Num. statements} \\   
    \hline
    Accessibility & 64\% & 160 \\
    Perceived trust & 59\% & 140 \\
    Human-likeness & 54\% & 100 \\
    Intention to use & 52\% & 80 \\
    Comprehensibility & 44\% & 160 \\
    Satisfaction & 42\% & 100 \\
    \hline
\end{tabular}
\end{center}
\caption{Positive perception on statements evaluating six characteristics of \texttt{Welzijn.AI}'s proof-of-concept interface (Figure \ref{fig:proof-of-concept_ex}). Num. statements refers to the total number of statements evaluated by 20 elderly individuals on a specific characteristic.}
\label{tab:user_eval}
\end{table}

\begin{table}[t] \small
\begin{center}
\begin{tabular}{c c c }
    \hline
    \textbf{Rank} & \textbf{Characteristic} & \textbf{Total points} \\   
    \hline
    1. & Responding empathetic & 101 \\
    2. & Exhibiting distinctive personality/character & 88 \\
    3. & Building a relation of trust & 86 \\
    4. & Developed social competencies & 82 \\ 
    5. & Establishing social relationships & 77  \\
    6. & Behaving transparantly & 65 \\
    7. & Using natural cues & 60 \\
    \hline
\end{tabular}
\end{center}
\caption{Ranking of desired social characteristics of Welzijn.AI by 20 elderly individuals.}
\label{tab:social_chars}
\end{table}

\subsubsection{Desired social characteristics}
\noindent The ranking of desired social characteristics for \texttt{Welzijn.AI} is given in Table \ref{tab:social_chars}. Being empathetic is for the elderly the most important characteristic, but they consider the use of natural cues least important. Looking at the three most important characteristics, elderly seem to have a preference for AI systems that can recognise and express emotions, that do not provide too generic or bland interaction but maintain some personality, and also for systems that instil a sense of trust. Elderly value this more than transparent model behaviour (e.g. the model being able to state what it does with the data), and showing natural behaviour (e.g. using fillers in interaction). 

\section{Discussion}\label{discussion}
\noindent This paper illustrated in what ways experts/stakeholders can be involved in the early phases of developing AI systems for elderly care, and presented insights gathered from three complementary stakeholder evaluations. In this section we place our findings in a broader context.

Though each expert session targeted different aspects of \texttt{Welzijn.AI}, they provided both shared and complementary insights. Non-elderly experts in the panel interviews (Section \ref{results:SWOT}) pointed out unclarity regarding the goal and broader utility of the application. Similarly, a majority of the elderly in the proof-of-concept evaluation (Section \ref{elderly_eval}) did not regard the application as comprehensible, for example did not grasp basic functionality of the technology as expressed by some icons. Experts in the co-creation session (Section \ref{results:co-creation}) provided possible solutions to this issue, by stressing value requirements that enable practice, education and support in the use of \texttt{Welzijn.AI}. Such suggestions bear on the \textit{context} of the technology instead of the technology itself, such as the existence of a help desk or demo session, which suggests AI systems for elderly care mandate a broader perspective on and longer-term investment in elderly environments. The emphasis on the context of new technology further resonated in the expert panel interviews, that highlighted the potential of \texttt{Welzijn.AI} to activate the social network of an elderly individual.

One strength of \texttt{Welzijn.AI} as pointed out in expert panel interviews was that it may help combat loneliness. We can complement this with results from the expert proof-of-concept evaluation, that identified priorities for chatbot design to help align chatbots with elderly preferences: they value empathy and personality/character in the conversations, more than information on transparency or sophisticated natural cues in interaction. Though the latter issues may be less important for the elderly, from the panel expert session and co-creation session we found that privacy as underlying value in various forms is for many non-elderly experts an important issue. This pertains for example to information access, continuous monitoring, but also safe data storage, and will likely remain important in a time where responsible AI becomes increasingly important, not only from an ethical but also from a legal perspective.

Elderly in the expert proof-of-concept evaluation agreed that \texttt{Welzijn.AI} is accessible, trustworthy, human-like, and instils an intention to use the system, but disagree that the system is comprehensible and satisfying. Yet, it is salient that the elderly were divided among most topics, as we found no large majorities that positively or negatively evaluate a design characteristic. This suggests that for AI applications for the elderly, there is still a mismatch between what developers and the elderly consider a comprehensible and satisfying app. Earlier work underscores the need to develop digital health applications from a user-centred perspective to avoid such mismatches \cite{mathews2019digital}. On the other hand it may signal a mismatch that cannot be easily resolved -- some elderly are tech-savy in that they use smartphones or have smart voice-controlled assistants at home, hence may feel disparaged when they are presented an application that was designed for elderly with zero experience with technology.     

Overall, our findings align with work suggesting that incorporating all stakeholder perspectives remains a key challenge for digital systems in health, since such perspectives align but also diverge as we saw in the co-creation session, but nevertheless critical to enhance application functionality and efficacy \cite{sedhom2021mobile, zainal2023usability}. Still, we believe our results to be helpful for developers and researchers working on AI systems in and beyond healthcare. For \texttt{Welzijn.AI} a working prototype primarily for demo purposes has been developed, that will be available online, but for which currently a demo video can be consulted, that may inspire future work on AI systems for elderly well-being.\footnote{The video (and later also application) can be found at \url{https://osf.io/ezyna/}.}

\section{Contributors}
\noindent Bram van Dijk contributed to the analysis and interpretation of the data, drafting the article and critical revision of the submitted manuscript. Armel Lefebvre contributed to the conception and design of the study, and critical revision of the submitted manuscript. Marco Spruit contributed to the conception and design of the study, acquisition of data, drafting the article, and reviewing the manuscript. All authors saw and approved the final version and no other person made a substantial contribution to the paper. 

\section{Funding}
\noindent The expert co-creation session took place within the project `Ethical appraisal of AI-driven MedTech for Trustworthy AI – development of a participatory, patient-centric methodology and toolkit', financed by ZonMw, grant number 10580012210019 of the program HTA Methodology 2021-2024.

\section{Ethical approval}
\noindent Apart from the considerations discussed in Section \ref{methods}, no separate ethical approval was obtained for the stakeholder evaluations in this study. 

\section{Data sharing and collaboration}
\noindent There are no linked research data sets for this paper, but datasets are available from the corresponding author upon request.

\section{Declaration of competing interest}
\noindent The authors declare that they have no conflicts of interests.

\section{Acknowledgments}\label{acknowledgments}
\noindent Many professionals and students have contributed already to the development of \texttt{Welzijn.AI}. For valuable input and/or work we thank Runda Wang and Casper van Wordragen (Leiden Institute of Advanced Computer Science), Danielle Matser and Vera Vroegop (Zorginstituut Nederland), Sytske van Bruggen and Anne-Wil Eewold (Hadoks), Els van de Boogart and Ildikó Vajda (Patiëntenfederatie Nederland), Mirjam de Haas and Melvin Alomerović (Hogeschool Utrecht), Wieke Hengeveld (MantelzorgNL), Nynke Slagboom and Hedwig Vos (Leiden University Medical Center), Edwin de Beurs (Leiden University), 20 elderly individuals from Dutch nursing homes in the South-West of the Netherlands, 15 Dutch AI Parade Dialogue participants, and 11 software engineering students (Leiden Institute of Advanced Computer Science). 

\bibliographystyle{elsarticle-num} 
\bibliography{natbib}





\section*{Supplementary material}

SUPPLEMENTARY MATERIALS FOR

# Welzijn.AI: Developing Responsible Conversational AI for Elderly Care through Stakeholder Involvement

Here we provide 1) all use cases from the expert co-creation session (Section 2.2.) and 2) the items used in the expert proof-of-concept evaluation (see Section 2.3.).

## Use cases (Section 2.2.)

**Use case 1, valued with 130 dollars, was created by the professional caregiver:**

*Ms. A lives alone in an appartment on the second floor. Normally she does a little stroll with her walker. Once every two days she does her grocery shopping. She manages to be away for about one hour. When she comes home she checks in with her robot. One day this happens much later, she has been away for two hours! She sat an hour in front of the stairs because she did not dare climbing the stairs. The robot asks what happened. Did Ms. fall? How did the walking go? After two to three hours, a warning signal is sent to a contact.*

**Use case 2, valued with 100 dollars, was created by informal caregiver 1:**

*Son (35 year old) lives in the North of the country, mother lives in the West. Mother is often alone. Son is worried about the risk of falling and her mental wellbeing. The robot keeps an eye on mother and signals if patterns change. Son receives a notification and calls mother to see how it is going.*

**Use case 3, valued with 60 dollars, was created by the developer:**

<Ad hoc chit chatting>
Mr. Doe:         Hi Wiebert… <touch>
Robot:           Hello mr. Doe, everything alright?
Mr. Doe:         It's quite something…
< Speech and language analysis>
Robot: I see the sun's out! Have you been outside yet?

## Items (Section 2.3.)

Here we show all items presented to the elderly, sorted by category. The survey was conducted with Dutch elderly individuals so questions are translations.

**Accessibility** (5-point Likert scale):
1. Text and icons are large enough
2. The 'settings' button is easy to spot
3. The functions of the icons on the screen are clear
4. The language used is easy to grasp
5. The contrasts between the background and text boxes is clear
6. I understand how I can respond with a message after the chatbot's introduction
7. I understand where I have to press when I want to send a message

8. I understand where I need to press when I want messages to be read out loud

**Comprehensibility** (5-point Likert scale):
1. The system looks needlessly complex
2. The system looks as if it is easy to use
3. I need the help of someone with technical knowledge if I want to use this system
4. I think I can master this system reasonably fast
5. I can see the point of this system
6. I think the system is hard to use
7. I think I can reliably use this system
8. I think that I will use this system frequently

**Intention to use** (5-point Likert scale):
1. I intend to use this system
2. I am not planning to use this system
3. I would only use this system if the GP recommends it
4. I would recommend this system to others

**Perceived trust** (5-point Likert scale):
1. I think that this system is reliable
2. I think that the technology is reliable in general
3. If this system is reliable, it will yield good results
4. I think this system is able to do what it says it will do
5. I do not trust this system
6. I think this system can address my needs when it comes to filling out surveys
7. I think that this system can work on its own without the assistance of humans
8. Overall I have trust in this system

**Satisfaction** (5-point semantic differential scale):
1. terrible – wonderful
2. easy – hard
3. frustrating – satisfying
4. satisfactory – not satisfactory
5. boring – fun

**Human-likeness** (5-point semantic differential scale):
1. fake – real
2. machinelike – humanlike
3. not conscious – conscious
4. artificial – natural


\end{document}